\documentclass[a4paper,12pt]{article}

\usepackage{amsmath,amssymb}
\usepackage{slashed}

\usepackage{cite}
\usepackage{color}
\usepackage{graphicx}

\makeatletter

\def\del{\partial}

\makeatletter
\g@addto@macro\bfseries{\boldmath}
\makeatother

\def\half{{\frac{1}{2}}}

\def\unit{{1\kern-.65ex {\rm l}}}
\def\1{{1\kern-.65ex {\rm l}}}

\def\be{\begin{equation}}
\def\ee{\end{equation}}

\newcount\hour \newcount\minute
\hour=\time \divide \hour by 60
\minute=\time
\def\now{%
\ifnum \hour<13
  \ifnum \hour=0 \advance \hour by 12 \number\hour:\else \number\hour:\fi%
     \ifnum \minute<10 0\fi%
     \number\minute%
\ A.M.%
\else \advance \hour by -12 \number\hour:%
  \ifnum \minute<10 0\fi%
  \number\minute%
  \ P.M.%
\fi%
}

\makeatother

\begin{document}

\baselineskip=18pt  
\numberwithin{equation}{section}  




\renewcommand{\thefootnote}{\fnsymbol{footnote}}
\setcounter{footnote}{1}
\thispagestyle{empty}

\vspace*{-2cm} 
\begin{flushright}
YITP-17-37
\end{flushright}

\vspace*{2.cm} 
\begin{center}
 {\LARGE Matching Renormalisation Schemes in Holography}\\
 \vspace*{1.7cm}
 {\large Shinji Hirano$^{1,2,}$\footnote{shinji.hirano@wits.ac.za}}\\
 \vspace*{1.0cm} 
{\it  $^1$ School of Physics and Mandelstam Institute for Theoretical Physics
\&
 DST-NRF Centre of Excellence in Mathematical and Statistical Sciences (CoE-MaSS),\\
  University of the Witwatersrand,\\
  WITS 2050, Johannesburg, South Africa\\[1ex]
  $^2$ Center for Gravitational Physics,
Yukawa Institute for Theoretical Physics,\\
Kyoto University, Kyoto 606-8502, Japan}

\end{center}
\vspace*{1.5cm}

\noindent
In holography there is a one-to-one correspondence between physical observables in the bulk and boundary theories.
To define physical observables, however, regularisation needs to be implemented in both sides of the correspondence. It is arguable whether the correspondence should extend to regularisation and renormalisation scheme which are not physical in the conventional sense. However, if we are to take the renormalisation group (RG) interpretation of holography seriously, its precise understanding appears to require the matching of regularisations and renormalisation schemes in the bulk and boundary theories. We address this question in the AdS$_5$/CFT$_4$ correspondence by considering a simplest physical quantity, the Casimir energy of the ${\cal N}=4$ super Yang-Mills (SYM) theory on $\mathbb{R}\times S^3$, in a $\zeta$-function regularisation and show that there are choices of scheme which match the bulk AdS result including the radial cutoff dependent corrections when the cutoff is kept finite. We further discuss the implication of this result to the RG interpretation of holography.

\newpage
\setcounter{page}{1} 
\renewcommand{\thefootnote}{\arabic{footnote}}
\setcounter{footnote}{0}





\section{Introduction}
\label{sec:Introduction}

Central to the idea of the AdS/CFT correspondence is the emergence of an extra dimension. 
The radial coordinate of the $(d+1)$-dimensional AdS space emerges as the energy scale of the $d$-dimensional field theory defined on the boundary \cite{Maldacena:1997re}. This, in particular, implies that moving from the boundary into the bulk of space may have an interpretation as the RG flow of the boundary field theory from UV to IR \cite{Balasubramanian:1999jd, deBoer:1999tgo}. In order to discuss the RG flow of the field theory, we need to introduce two scales, the UV cutoff $\Lambda$ to regularise the theory and the IR scale $\Lambda_{\rm IR}$ at which the low energy effective theory is defined. 
These two scales are related to radial scales, i.e. constant radial slices, of the (asymptotically) AdS space. 
The scheme of renormalisation, meanwhile, specifies how to remove contributions which diverge as the UV cutoff $\Lambda$ is sent to infinity. In particular, in the exact RG approach the choice of regularisation and renormalisation scheme is fixed by the choice of cutoff functions for propagators, commonly denoted by $K(p/\Lambda)$
\cite{Wilson, Polchinski, Niigata}\footnote{From section \ref{sec:RenSchemHolo} onwards we use the symbol $\eta$ instead of $K$ for cutoff functions. }
On the gravity side the standard renormalisation scheme is fixed by the choice of covariant counter terms on the cutoff radial slice added to the gravity action\cite{deHaro:2000vlm, Emparan:1999pm}.

In holography there is a one-to-one correspondence between physical observables in the bulk and boundary theories. Although it is arguable whether the correspondence should extend to regularisation and renormalisation scheme which are not physical in the conventional sense, if we are to take the RG interpretation of holography seriously, its precise understanding appears to require the matching of regularisations and renormalisation schemes in the bulk and boundary theories. This is the question we wish to address and elucidate in this paper. More concretely, we study how precisely the cutoff or the scale $\Lambda$ (and $\Lambda_{\rm IR}$) of the field theory is related to the radial scale in AdS and whether and how the cutoff function $K(p/\Lambda)$, properly adapted to our specific context, can be determined to reproduce the result obtained by the holographic renormalisation scheme of \cite{deHaro:2000vlm, Emparan:1999pm} when the cutoff is kept finite. 
As a simplest workable example, we consider the AdS$_5$/CFT$_4$ correspondence and, in particular, the Casimir energy of the ${\cal N}=4$ SYM theory on $\mathbb{R}\times S^3$ in a $\zeta$-function regularisation and illustrate how the above matching can be done operationally in detail.
We further discuss Wilsonian RG flow in a similar spirit to \cite{Heemskerk:2010hk, Faulkner:2010jy}, working in this same example, and elaborate on how integrating-out of high momentum modes is implemented on both sides of the correspondence.

The organisation of the paper is as follows: In section \ref{sec:RenSchemHolo}, as just mentioned, we study and review the Casimir energy of the ${\cal N}=4$ SYM theory on $\mathbb{R}\times S^3$ on both sides of the correspondence and present the results including the cutoff/radial-dependent corrections with particular emphasis on the regularisations and the choice of renormalisation schemes.
In section \ref{sec:Matching} we ask whether and how the results on the two sides can match and show that there exist such choices of the cutoff function that yield the matching of the Casimir energies with the cutoff/radial-dependent corrections. 
In section \ref{sec:RGflow} we discuss Wilsonian RG in holography and how it can be understood in light of the results we find in section \ref{sec:RenSchemHolo} and \ref{sec:Matching}. We close this paper with a summary of the results and short discussions in section \ref{sec:discussions}.

\section{Renormalisation schemes in holography}
\label{sec:RenSchemHolo}

The standard holographic renormalisation scheme \cite{deHaro:2000vlm, Emparan:1999pm} proceeds in two steps: (1) Regularising the theory by introducing an IR (in the sense of bulk and corresponding to UV in the field theory) cutoff $r_{\Lambda}$ of the radial coordinate $r$ near the boundary and (2) adding the covariant counter-term action at the cutoff surface $r=r_{\Lambda}$ so that divergences which arise as $r_{\Lambda}\to\infty$ are eliminated. For a given quantity of interest, this scheme yields a finite piece plus cutoff-dependent corrections that vanish as $r_{\Lambda}\to\infty$. Although the finite piece is of primary interest and one is mostly interested in whether it can be matched to the dual field theory result, it is the interest of this paper if the cutoff-dependent corrections too can match the field theory result. 

To address this question in a simplest setting, we consider the Casimir energy of the ${\cal N}=4$ SYM on $\mathbb{R}\times S^3$ which has been known to agree exactly with the holographic calculation \cite{Balasubramanian:1999re}.
In the gravity description the Casimir energy corresponds to the on-shell free energy evaluated on the global AdS$_5$ space without any other fields turned on. Here we recall the holographic renormalisation scheme for the Casimir energy reviewed in \cite{Marino:2011nm} and elaborate on the $\zeta$-function regularisation adopted in the field theory calculation.

\subsection{The gravity description}
\label{sec:gravdescription}

The gravity action $I$ consists of three parts, namely (1) the bulk Einstein-Hibert action $I_{\rm bulk}$, (2) the boundary Gibbons-Hawking term $I_{\rm GH}$ \cite{Gibbons:1976ue} and (3) the covariant counter-term $I_{\rm ct}$ on the cutoff boundary surface at $r=r_{\Lambda}$ \cite{deHaro:2000vlm, Emparan:1999pm}:
\be
I=I_{\rm bulk}+I_{\rm GH}+I_{\rm ct}\ .
\label{action}
\ee
Each contribution has the following form,
\begin{align}
I_{\rm bulk}&=-{1\over 16\pi G_N}\int_{M} d^5x\sqrt{g}\left(R+{24\over L^2}\right)\ ,\label{bulkaction}\\
I_{\rm GH}&=-{1\over 8\pi G_N}\int_{\del M}d^4x\sqrt{\gamma}K\ ,\label{GHterm}\\
I_{\rm ct}&={1\over 8\pi G_N}\int_{\del M}d^4x\sqrt{\gamma}\left({3\over L}+{L\over 4}R(\gamma)\right)\ ,
\label{ctaction}
\end{align}
where $G_N$ is Newton's constant, $M$ the bulk spacetime, $\del M$ the boundary created by the IR (in the sense of bulk) cutoff at $r=r_{\Lambda}$, $L$ the AdS$_5$ radius, $\gamma$ the induced metric on $\del M$, and $K$ the extrinsic curvature of the boundary $\del M$. Note that higher order curvature terms could have been added to the counter-term \eqref{ctaction} and their absence corresponds to one specific choice of the renormalisation scheme.

The renormalised Casimir energy $E$ is the free energy obtained from the on-shell gravity action $I$ on the Euclidean global AdS$_5$ metric
\begin{align}
ds^2=V(r)d\tau^2+{dr^2\over V(r)}+r^2d\Omega_3^2
\qquad\mbox{with}\qquad V(r)=1+{r^2\over L^2}
\end{align}
where the Euclidean time $\tau$ has the periodicity $\beta$ and $d\Omega_3^2$ is the metric on $S^3$. The on-shell value of each part of the gravity action is, respectively, given by
\begin{align}
I_{\rm bulk}&={\beta\over 8\pi G_N L^2}V_{S^3}\, r_{\Lambda}^4\ ,\\
I_{\rm GH}&=-{\beta\over 8\pi G_N}V_{S^3}\, r_{\Lambda}^2\left({3\over r_{\Lambda}}V(r_{\Lambda})
+\half V'(r_{\Lambda})\right)\label{GHvalue}\ ,\\
I_{\rm ct}&={\beta\over 8\pi G_N}V_{S^3}\, r_{\Lambda}^3\left({3\over L}+{3L\over 2r_{\Lambda}^2}\right)V(r_{\Lambda})^{\half}\ ,\label{ctvalue}
\end{align}
where the dash in $V'(r_{\Lambda})$ denotes the derivative w.r.t. $r$ and the volume of the three-sphere $V_{S^3}=2\pi^2$. Note that as $r_{\Lambda}\to\infty$, the first two, $I_{\rm bulk}$ and $I_{\rm GH}$, only have divergent contributions which are cancelled by those in the counter-term $I_{\rm ct}$, and the finite contribution, i.e., $\beta E$, solely comes from $I_{\rm ct}$.
The on-shell gravity action, including subleading corrections that vanish as $r_{\Lambda}\to\infty$, yields
\begin{align}
I=\beta{3\pi L^2\over 32 G_N}\left[1+\sum_{n=1}^{\infty}\left({L^2\over 4r_{\Lambda}^2}\right)^n
\left\{{(-1)^n(2n+2)!\over(2n+1)((n+1)!)^2}-{(-1)^n(2n+4)!\over 2(2n+3)((n+2)!)^2}\right\}\right]\ .
\end{align}
The AdS/CFT correspondence relates Newton's constant $G_N$ to the rank $N$ of the ${\cal N}=4$ SYM by $N^2=\pi L^3/(2G_N)$. The renormalised Casimir energy $E_{\Lambda}=\beta^{-1}I$ at finite cutoff $r_{\Lambda}$ is thus found to be
\begin{align}
E_{\Lambda}={3N^2\over 16L}\left[1+\sum_{n=1}^{\infty}\left({L^2\over 4r_{\Lambda}^2}\right)^n
\left\{{(-1)^n(2n+2)!\over(2n+1)((n+1)!)^2}-{(-1)^n(2n+4)!\over 2(2n+3)((n+2)!)^2}\right\}\right]\ ,
\label{CasimirGrav}
\end{align}
where the first term ${3N^2\over 16L}$ is the physical Casimir energy $E=\lim_{r_{\Lambda}\to\infty}E_{\Lambda}$.

\subsection{The field theory description}
\label{sec:FTdescription}

As mentioned earlier, an exact agreement has been found between the Casimir energy of the ${\cal N}=4$ SYM on $\mathbb{R}\times S^3$ and that of the gravity dual \cite{Balasubramanian:1999re}. It is worth emphasing that the field theory result was obtained in the weak coupling (free) limit, which indicates that the Casimir energy is protected from quantum corrections. Alternatively, this non-renormalisation property may have been anticipated from the fact that the strong coupling result $E={3N^2\over 16L}$ in \eqref{CasimirGrav} has no dependence on the 't Hooft coupling $\lambda=g_{\rm YM}^2N$. 
In principle, the field theory calculation can be done directly at strong couplings using the localisation technique \cite{Witten:1988ze,Pestun:2007rz} and must yield the same result as the weak coupling one. However, there is a caveat: As studied and discussed in \cite{Assel:2014paa, Lorenzen:2014pna, Assel:2015nca}, the localisation computation yields a different Casimir energy, dubbed supersymmetric Casimir energy. As clarified in \cite{Lorenzen:2014pna}, by performing a large gauge transformation to the background gauge field, the supersymmetric Casimir energy can be interpolated to the standard Casimir energy, but the resulting shift to the background gauge field does not respect the periodicity of the Euclidean time, leaving a technical issue in the localisation calculation.

It should also be noted that there is, in general, the issue of gauge invariant regularisation. 
Except for the dimensional regularisation which does not require an explicit cutoff, most regularisations break gauge invariance by the introduction of the cutoff. This issue is virtually circumvented in the $\zeta$-function regularisation of the Casimir energy.

The Casimir energy of the ${\cal N}=4$ SYM on $\mathbb{R}\times S^3$ is the sum of zero point energies of the six scalars $\Phi$, four fermions $\Psi$ and one vector $A$:
\begin{align}
E=N^2\left(6E_{\Phi}+4E_{\Psi}+E_{A}\right)\ ,
\label{bareCasimir}
\end{align} 
where each zero point energy is an infinite sum over modes
\begin{align}
E_{\Phi}&={1\over 2L}\sum_{m=1}^{\infty}{m^2\over m^s}={1\over 2L}\sum_{n=1}^{\infty}n^{2-s}\ ,\label{EPhi}\\
E_{\Psi}&=-{1\over 2L}\sum_{m=1}^{\infty}{2m(m+1)\over \left(m+\half\right)^s}
=-{1\over L}\sum_{n=1}^{\infty}\left[(2^{s-2}-1)n^{2-s}-{1\over 4}(2^{s}-1)n^{-s}\right]\ ,\label{EPsi}\\
E_{A}&={1\over 2L}\sum_{m=1}^{\infty}{2m(m+2)\over (m+1)^s}={1\over L}\sum_{n=1}^{\infty}(n^{2-s}-n^{-s})\ ,
\label{EA}
\end{align}
where $s=-1$ and the parameter $s$ is introduced in anticipation of the $\zeta$-function regularisation. However, instead of simply analytically continuing $\sum_{n=1}^{\infty}n^{-s}$ to $\zeta(s)$, we employ a more elementary method of renormalisation: regularising the infinite sum by introducing cutoff functions and appropriately choosing them so that the divergences, as the cutoff $\Lambda \to \infty$, are eliminated.

The regularised Casimir energy, which is the field theory counterpart of \eqref{CasimirGrav} , takes the form \cite{TerryTao}
\begin{align}
E_{\Lambda}={3N^2\over 2L}\left[5\sum_{n=1}^{\infty}n^3\,\eta_3(n/\Lambda)-\sum_{n=1}^{\infty}n\,\eta_1(n/\Lambda)\right]\ ,\label{regCasimir}
\end{align}
where the cutoff functions $\eta_1(x)$ and $\eta_3(x)$ must at least satisfy the following properties:
\be
\eta_s(0)=1\qquad\mbox{and}\qquad\eta_s(x)\stackrel{x\to\infty}{\longrightarrow} 0\quad(\mbox{faster\,\,than}\,\, x^{-s-1})\ .
\ee
This is the regularised form of \eqref{bareCasimir}. We now recall the Euler-Maclaurin formula
\begin{align}
\sum_{n=1}^{\infty}n^s\,\eta_s(n/\Lambda)=-{B_{s+1}\over s+1}+\Lambda^{s+1}\int_0^{\infty}dx\, x^s\eta_s(x)
-\!\!\!\!\sum_{k=\left[{s+1\over 2}\right]+1}^{\infty}{B_{2k}\over (2k)!}{s!\over \Lambda^{2k-1-s}}\eta^{(2k-1-s)}_s(0)\ ,
\label{EMexpansion}
\end{align}
where $B_l$ is the $l$-th Bernoulli number and $\eta_s^{(m)}(0)$ is the $m$-th derivative of the cutoff function $\eta_s(x)$ at $x=0$. This has to be understood as an asymptotic expansion. 
Using this formula, the regularised Casimir energy \eqref{regCasimir} yieds
\begin{align}
E_{\Lambda}=E_{\Lambda}^{(\rm div)}+E+E_{\Lambda}^{(\rm sub)}
\label{CFTCasimir}
\end{align}
where each contribution is given by
\begin{align}
E_{\Lambda}^{(\rm div)}&={3N^2\over 2L}\left[5\Lambda^3\int_0^{\infty}dx\, x^3\eta_3(x)+\Lambda\int_0^{\infty}dx\, x\eta_1(x)\right]\ ,\\
E&=-{3N^2\over 2L}\left({5\over 4}B_4-{B_2\over 2}\right)={3N^2\over 16L}\ ,\\
E_{\Lambda}^{(\rm sub)}&={3N^2\over 2L}\sum_{k=1}^{\infty}{1\over \Lambda^{2k}}\left[{B_{2k+2}\over(2k+2)!}\eta_1^{(2k)}(0)-{30B_{2k+4}\over(2k+4)!}\eta_3^{(2k)}(0)\right]\ .
\end{align}
Note that the cutoff independent finite piece $E$ is the same physical Casimir energy as the one calculated by the standard $\zeta$-function regularisation which agrees with the gravity result. 
To renormalise the regularised Casimir energy, {\it i.e.} to render $E_{\Lambda}^{(\rm div)}=0$, 
the following conditions must be imposed on the cutoff functions:
\begin{align}
\int_0^{\infty}dx\, x^3\eta_3(x)=\int_0^{\infty}dx\, x\eta_1(x)=0\ .
\label{nodiv}
\end{align}
The renormalisation scheme is determined by the choice of the cutoff functions $\eta_1(x)$ and $\eta_3(x)$.

\section{Matching the renormalisation schemes}
\label{sec:Matching}

We now ask whether there exists a choice of the renormalisation scheme of the field theory that can be matched to the holographic one. In other words, we look for a choice of the cutoff functions $\eta_1(x)$ and $\eta_3(x)$ for which the  Casimir energy \eqref{CFTCasimir} of the field theory agrees with that of the gravity dual \eqref{CasimirGrav} including the cutoff dependent subleading corrections. 

There may be no reason for the regularised Casimir energy to be protected from $\lambda$-dependent quantum corrections. Thus it might not be sensible to ask if the renormalisation scheme at weak couplings matches the one at strong couplings. However,  provided that the aforementioned technical issue can be overcome, it seems plausible to assume that the localisation calculation performed directly at strong couplings gives rise to the same Casimir energy as the one in the free limit \eqref{bareCasimir} prior to the regularisation. We shall thus work under this assumption.

As the finite piece $E$ has been already matched, we are left with the matching of the subleading corrections: 
\begin{align}
&\left({L^2\over 4r_{\Lambda}^2}\right)^k
\left[{(-1)^k(2k+2)!\over(2k+1)((k+1)!)^2}-{(-1)^k(2k+4)!\over 2(2k+3)((k+2)!)^2}\right]\label{matching}\\
&\hspace{2cm}={8\over \Lambda^{2k}}\left[{B_{2k+2}\over(2k+2)!}\eta_1^{(2k)}(0)-{30B_{2k+4}\over(2k+4)!}\eta_3^{(2k)}(0)\right]\nonumber
\end{align}
for $k\ge 1$. It is apparent that the solution exists but is not unique.\footnote{Non-uniqueness is not an issue, rather it can be interpreted as spurious scheme-independence.} {\it A priori}, the cutoff functions $\eta_1(x)$ and $\eta_3(x)$ can be independent, and thus this matching condition only determines $\eta_3^{(2k)}(0)$ in terms of $\eta_1^{(2k)}(0)$, or {\it vice versa}.
However, since the scalars $\Phi$, the fermions $\Psi$, and the vector $A$ are all related by supersymmetries, it may make more sense to relate these cutoff functions for the Casimir energies \eqref{EPhi}--\eqref{EA} accordingly. 
Here we assume a particular relation between them which appears to be minimal and natural:
\be
\eta_1(x)=\eta(x)\qquad\mbox{and}\qquad
\eta_3(x)={\eta^{(2)}(x)\over\eta^{(2)}(0)}\qquad\mbox{with}\qquad\eta(0)=1\ .\label{eta1eta3}
\ee
With this choice the second equality of \eqref{nodiv} for the absence of divergences guarantees the first, provided that $\eta(x)$ falls off faster than $x^{-2}$ as $x\to\infty$. The matching condition \eqref{matching} then becomes a recursion relation for $\eta^{(2k)}(0)$.
The general solution can be found, but it is still not unique as $\eta^{(2)}(0)$ can be freely chosen. Here we only present a special simplest solution.
It can be found by exploiting the invariance of  \eqref{matching} under the scaling and the shift
\begin{align}
\Lambda&\to c\Lambda\equiv\widetilde{\Lambda}\ ,\nonumber\\
\eta_1^{(2k)}(0)&\to  c^{2k}\left[\eta_1^{(2k)}(0)+a{(2k+2)!\over B_{2k+2}}\right]\equiv \tilde{\eta}_1^{(2k)}(0)\ ,\\
\eta_3^{(2k)}(0)&\to c^{2k}\left[\eta_3^{(2k)}(0)+a{(2k+4)!\over 30 B_{2k+4}}\right]\equiv \tilde{\eta}_3^{(2k)}(0)\ .\nonumber
\end{align}
for any constants $a$ and $c$.
In the matching condition \eqref{matching} we identify 
\begin{align}
\widetilde{\Lambda}&={2r_{\Lambda}\over L}\ ,\nonumber\\
{B_{2k+2}\over(2k+2)!}\tilde{\eta}_1^{(2k)}(0)&={(-1)^k(2k+2)!\over 8(2k+1)((k+1)!)^2}\ ,\\
{30B_{2k+4}\over(2k+4)!}\tilde{\eta}_3^{(2k)}(0)&={(-1)^k(2k+4)!\over 16(2k+3)((k+2)!)^2}\ ,\nonumber
\end{align}
where the constants $a$ and $c$ are determined by requiring the relation \eqref{eta1eta3}. They are found to be
\be
a={1\over 6}\qquad \mbox{and}\qquad c^2=-2\ ,
\ee
which in turn yields
\begin{align}
\eta^{(2k)}(0)={(2k+2)!\over B_{2k+2}}\left[-{1\over 6}+{(2k)!\over 2^{k+2}k!(k+1)!}\right]\ .
\label{etaderivatives}
\end{align}
Note that although this result is found for $k\ge 1$, we can sensibly extend it to $k=0$ since it so happens that $\eta(0)\equiv\eta^{(0)}(0)=1$, satisfying the boundary condition in \eqref{eta1eta3}. This turns out to be a rather important property, as this justifies the analytic continuation we perform later in \eqref{eta_analytic}.
Meanwhile, the cutoff $\Lambda$ as defined in \eqref{regCasimir} has to be pure imaginary since so is the constant $c$ and $c\Lambda=2r_{\Lambda}/L$ is real. However, the cutoff function $\eta(n/\Lambda)$ is real as implied by the Euler-Maclaurin expansion \eqref{EMexpansion} with $s=1$, and thus $\Lambda$ being pure imaginary is not an issue. 
As a consequence of the matching, we find that the UV cutoff $|\Lambda|$ in the field theory is related to the IR cutoff $r_{\Lambda}$ in gravity by
\be
|\Lambda|={\sqrt{2}r_{\Lambda}\over L}\ .
\label{cutoffmatch}
\ee

We now wish to fully construct the cutoff function $\eta(x)$ from the data \eqref{etaderivatives}. For this purpose it is most convenient to consider the Mellin transform defined by
\be
{\cal M}\left[\eta(x); s\right]\equiv \eta^{\ast}(s)=\int_0^{\infty}dx\,\eta(x) x^{s-1}
\ee
where the integral is well-defined in the range $0<\mbox{Re}(s)<M$, provided that $\eta(x)\to{\cal O}(x^{-M})$ as $x\to\infty$. The Mellin transform is invertible and the inverse transform is given by
\be
\eta(x)={1\over 2\pi i}\int_{c-i\infty}^{c+i\infty}ds\,\eta^{\ast}(s) x^{-s}
\ee
where $0<c<M$ and $\eta^{\ast}(s)$ must fall off fast enough as $s\to\pm i\infty$. For $x\ll 1$, this tells us that
\be
\eta(x)=\sum_{m=0}^{\infty}\left[{\rm Res}\,\eta^{\ast}(s)\right]_{s=-m}x^m
\quad\Longrightarrow\quad
\eta^{(2k)}(0)=(2k)!\left[{\rm Res}\,\eta^{\ast}(s)\right]_{s=-2k}\ .
\label{residue}
\ee
We also have the condition for the absence of divergences, which is given by
\be
\int_0^{\infty}dx\,\eta(x)x=0\quad\Longrightarrow\quad \eta^{\ast}(s=2)=0
\label{nodivergence}
\ee
which is to  ensure $M>2$ so that $\eta(x)\to{\cal O}(x^{-2})$ as $x\to\infty$.

Thus the Mellin transform $\eta^{\ast}(s)$ must have simple poles at $s=-2k$ ($k\in\mathbb{Z}_{\ge 0}$) and a zero at $s=2$. The choice of $\eta^{\ast}(s)$ which satisfies \eqref{residue} and \eqref{nodivergence} is not unique.
Here we provide a particular choice to show that such $\eta^{\ast}(s)$ exists:
\begin{align}
\eta^{\ast}(s)={\eta^{(-s)}(0)\over\Gamma(1-s)}\left[{\pi^2\over 4\sin^2\left({\pi s\over 2}\right)}{\zeta'(s-2)\over\zeta(s-2)}
\left({\zeta(s-4)\over\zeta'(s-4)}\right)^2\right]
\label{etastar}
\end{align}
where the Riemann $\zeta$-function $\zeta(z)$ has zeros at $z=-2, -4, -6, \cdots$, besides nontrivial zeros along the imaginary axis at ${\rm Re}(z)=\half$, and a pole at $z=1$, and $\eta^{(-s)}(0)$ is an analytic continuation of the $2k$-th derivative $\eta^{(2k)}(0)$ in \eqref{etaderivatives} and defined by
\begin{align}
\eta^{(-s)}(0)={(2\pi)^{2-s}e^{{i\pi s\over 2}}\over 2 \zeta(2-s)}\left[-{1\over 6}
+{\Gamma\left(\half-{s\over 2}\right)\over 2^{{s\over 2}+2}\sqrt{\pi}\Gamma\left(2-{s\over 2}\right)}\right]\ ,
\label{eta_analytic}
\end{align}
where we used $\Gamma\left(k+\half\right)=\sqrt{\pi}{(2k)!\over 2^{2k}k!}$ and 
$
{B_{2k+2}\over (2k+2)!}=2(-)^{k}{\zeta(2k+2)\over(2\pi)^{2k+2}}\stackrel{2k\to -s}
{\longrightarrow}2e^{-{i\pi s\over 2}}(2\pi)^{s-2}\zeta(2-s)\ .
$
Note that the $\eta^{\ast}(s)$ as chosen in \eqref{etastar} falls off exponentially as  $s\to\pm i\infty$ due to the factor $1/\sin^2\left({\pi s\over 2}\right)$, satisfying the asymptotic condition at $s=\pm i\infty$.

To summarise this section, there exists, though not unique, a choice of the renormalisation scheme of the field theory that can be matched to the holographic one. The cutoffs are related by \eqref{cutoffmatch}, and the Mellin transform $\eta^{\ast}(s)$ of the cutoff function $\eta(x)$ can be chosen to be \eqref{etastar} supplemented with \eqref{eta_analytic}.
The non-uniqueness of the choice of the cutoff functions is not an issue because it simply implies some degrees of scheme-independence left within this scheme.

\section{The RG flow}
\label{sec:RGflow}
 
In light of the matching between the holographic and the field theory renormalisation schemes studied in the previous sections, we shall discuss the holographic RG flow and its relation to the RG flow in the field theory.

To consider the RG flow in the field theory, one first needs to regularise the theory by introducing the UV cutoff $\Lambda$ and then integrate out momentum modes down to an IR scale $\Lambda_{\rm IR}$. In the case of the Casimir energy, this corresponds to considering the subtraction
\begin{align}
E_{\rm IR}\equiv E_{\Lambda}-E_{\Lambda_{\rm IR}}
={3N^2\over 2L}\sum_{n=1}^{\infty}\left[5n^3\left\{\eta_3\left({n\over \Lambda}\right)-\eta_3\left({n\over\Lambda_{\rm IR}}\right)\right\}-n\left\{\eta_1\left({n\over\Lambda}\right)-\eta_1\left({n\over\Lambda_{\rm IR}}\right)\right\}\right]
\label{FTRGflow}
\end{align}
since the cutoff functions in effect truncate the modes $n$ higher than the scale $\Lambda$ and thus the difference $E_{\rm IR}$ is a sum over the modes in the range $\Lambda_{\rm IR}\le n\le \Lambda$. In other words, the modes $n$ are integrated from the scale $\Lambda$ down to $\Lambda_{\rm IR}$ in $E_{\rm IR}$, whereas the IR modes lower than the scale $\Lambda_{\rm IR}$ are left intact. Note that as the IR cutoff $\Lambda_{\rm IR}\to 0$, the cutoff functions $\eta_1(n/\Lambda_{\rm IR})$ and  $\eta_3(n/\Lambda_{\rm IR})\to 0$ and all modes below the UV cutoff $\Lambda$ are integrated out. 

The discussions in the previous sections then suggest that the holographic RG flow of the Casimir energy is given by
\begin{align}
E_{\rm IR}= \beta^{-1}\left(I[r_{\Lambda}]-I[r_{\Lambda_{\rm IR}}]\right)
\end{align}
where $I[r_{\Lambda}]$ and $I[r_{\Lambda_{\rm IR}}]$ are the gravity actions with the boundaries at $r=r_{\Lambda}$ and $r=r_{\Lambda_{\rm IR}}$, respectively. To understand its connection to the RG flow, we recall that
\begin{align}
E_{\rm IR}=\beta^{-1}\left[\int_{r_{\Lambda_{\rm IR}}}^{r_{\Lambda}}\!\!drL_{\rm bulk}[g_{\mu\nu}^{\rm AdS}]
+I_{\rm GH}(r_{\Lambda})+I_{\rm ct}(r_{\Lambda})-I_{\rm GH}(r_{\Lambda_{\rm IR}})-I_{\rm ct}(r_{\Lambda_{\rm IR}})\right]
\label{GRRGflow}
\end{align}
where $L_{\rm bulk}[g_{\mu\nu}^{\rm AdS}]$, with $g_{\mu\nu}^{\rm AdS}$ being the metric on the global AdS space, is the on-shell bulk Lagrangian. This can be interpreted as integrating out the bulk fields from the radial scale $r_{\Lambda}$ down to $r_{\Lambda_{\rm IR}}$: At large $N$ which is the classical limit of gravity, path integrals over the bulk fields can be approximated by the saddle point. In the case of the Casimir energy, the saddle point is at $g_{\mu\nu}=g_{\mu\nu}^{\rm AdS}$ and all other fields collectively denoted by $\phi=0$. 
Thus, when integrating out, only the metrics $g_{\mu\nu}$ in the range $r_{\Lambda_{\rm IR}}\le r\le r_{\Lambda}$ take the saddle point value $g_{\mu\nu}^{\rm AdS}$, whereas those in the range $0\le r< r_{\Lambda_{\rm IR}}$ are intact, leaving path integrals over the bulk fields in the region $0\le r\le r_{\Lambda_{\rm IR}}$ unperformed.
Note that in accordance with the field theory case, as $r_{\Lambda_{\rm IR}}\to 0$, the boundary terms $I_{\rm GH}(r_{\Lambda_{\rm IR}})$ and $I_{\rm ct}(r_{\Lambda_{\rm IR}})\to 0$ as in \eqref{GHvalue} and \eqref{ctvalue} and all bulk fields are integrated out. 

With the choice of the cutoff functions as discussed in the previous section, this provides a simplest example of the matching between the holographic RG flow \eqref{GRRGflow}  and that of the field theory \eqref{FTRGflow}.

\section{Conclusions and discussions}
\label{sec:discussions}

Motivated by the RG interpretation of holography, we asked if and how regularisations and renormalisation schemes can be matched across two sides of the AdS/CFT correspondence, even though these are not physical observables in the conventional sense. To set the question concrete and sharp, we considered a simplest physical quantity, the Casimir energy of the ${\cal N}=4$ super Yang-Mills (SYM) theory on $\mathbb{R}\times S^3$, in a $\zeta$-function regularisation and showed that there is in fact a choice of scheme which matches the bulk AdS result including the radial cutoff dependence. 
It is worth noting that there are rather large degrees of ambiguities in the choice of scheme that can be matched to the holographic result. However, this non-uniqueness is not an issue because it simply implies some degrees of spurious scheme-independence left within this scheme.
It would be interesting to study if the matching of renormalisation schemes can be done for other quantities such as Wilson loops, correlation functions as well as the free energy, for example, of the ABJM theory for which the localisation provides direct strong coupling results. 

We also discussed the holographic RG flow in light of the $\zeta$-function regularisation and our choice of renormalisation scheme. There have been a number of works on the Wilsonian interpretation of the holographic RG flow, and our discussion was in line with \cite{Heemskerk:2010hk, Faulkner:2010jy}. We believe that, though modest, it is a step forward in that we made a concrete comparison between the holographic and Wilsonian RG flows, albeit in a simplest setting. The only bulk field involved was the metric $g_{\mu\nu}$, as we concerned ourselves with the vacuum of the field theory. In order to make more direct contact with the proposals in \cite{Heemskerk:2010hk, Faulkner:2010jy}, we need to consider correlation functions to include other bulk fields $\phi$ and their dual operators in the RG flow. We leave these studies for future works.

\section*{Acknowledgments}

We would like to thank Robert de Mello Koch, Daniel Elander, Sanefumi Moriyama, Tadakatsu Sakai and Masaki Shigemori
for useful discussions
and the Graduate School of Mathematics, Nagoya university, and the Yukawa Institute for Theoretical Physics for their hospitality. We would also like to thank Marcos Mari\~no whose series of lectures on localisation and matrix models at Nagoya university in 2013 inspired this work.
This work was supported in part by the National Research Foundation of South Africa and DST-NRF Centre of Excellence in Mathematical and Statistical Sciences (CoE-MaSS).
Opinions expressed and conclusions arrived at are those of the author and are not necessarily to be attributed to the NRF or the CoE-MaSS.

\if{
\appendix

\section{Formal expansion}
\label{app:expand_S}
}\fi


\end{document}